**Title**: Constraints on Climate and Habitability for Earth-like Exoplanets Determined from a General Circulation Model.

**Short Title:** Constraints on Climate and Habitability

**Authors:** Eric T. Wolf[1], Aomawa L. Shields [2,3,4,7], Ravi K. Kopparapu[5,6,7,8], Jacob Haqq-Misra[7,8], Owen B. Toon[1]

**Affiliations:**
[1]Laboratory for Atmospheric and Space Physics, Department of Atmospheric and Oceanic Sciences University of Colorado, Boulder, Colorado, USA.

[2]University of California, Irvine, Department of Physics and Astronomy, 4129 Frederick Reines Hall, Irvine CA 92697, USA

[3]University of California, Los Angeles, Department of Physics and Astronomy, Box 951547, Los Angeles, CA 90095, USA

[4]Harvard-Smithsonian Center for Astrophysics, 60 Garden Street, Cambridge, MA 02138, USA

[5]NASA Goddard Space Flight Center

[6]Department of Astronomy, University of Maryland

[7]NASA Astrobiology Institute's Virtual Planetary Laboratory, P.O. Box 351580, Seattle, WA 98195, USA

[8]Blue Marble Space Institute of Science, 1001 4th Ave, Suite 3201, Seattle, Washington 98154, USA

**Corresponding Author Information:**
Eric Wolf
Laboratory for Atmospheric and Space Physics
3665 Discovery Drive
Campus Box 600
University of Colorado
Boulder, CO 80303-7820
eric.wolf@colorado.edu
240-461-8336

**Abstract**

Conventional definitions of habitability require abundant liquid surface water to exist continuously over geologic timescales. Water in each of its thermodynamic phases interacts with solar and thermal radiation and is the cause for strong climatic feedbacks. Thus, assessments of the habitable zone require models to include a complete treatment of the hydrological cycle over geologic time. Here, we use the Community Atmosphere Model from the National Center for Atmospheric Research to study the evolution of climate for an Earth-like planet at constant $CO_2$, under a wide range of stellar fluxes from F-, G-, and K-dwarf main sequence stars. Around each star we find four stable climate states defined by mutually exclusive global mean surface temperatures ($T_s$); snowball ($T_s \leq 235$ K), waterbelt (235 K $\leq T_s \leq$ 250 K), temperate (275 K $\leq T_s \leq$ 315 K), and moist greenhouse ($T_s \geq 330$ K). Each is separated by abrupt climatic transitions. Waterbelt, temperate, and cooler moist greenhouse climates can maintain open-ocean against both sea-ice albedo and hydrogen escape processes respectively, and thus constitute habitable worlds. We consider the warmest possible habitable planet as having $T_s \sim 355$ K, at which point diffusion limited water-loss could remove an Earth ocean in ~1 Gyr. Without long timescale regulation of non-condensable greenhouse species at Earth-like temperatures and pressures, such as $CO_2$, habitability can be maintained for an upper limit of ~2.2, ~2.4 and ~4.7 Gyr around F-, G- and K-dwarf stars respectively due to main sequence brightening.

## 1. Introduction:

Detecting Earth-like extrasolar planets is one of the primary objectives of ongoing and future exoplanetary observational surveys. Upcoming missions like the *James Webb Space Telescope* (Gardner et al. 2006), the *Transiting Exoplanet Survey Satellite* (Ricker et al. 2014), and others, will greatly improve our ability to detect and then begin characterizing terrestrial planets in the habitable zones of other stars. Still, observations will remain sparse compared to solar system objects and thus climate models are needed to interpret and understand these remote data. The conventional definition of the habitable zone requires liquid water to be extant and abundant at the surface continuously for at least several billion years in order for advanced life to evolve (Hart 1979). By definition, the fate of habitable worlds is inextricably tied to water and its associated feedbacks on the climate system. Thus, at its heart, the study of the habitable zone for Earth-like planets is the study of the fundamental evolutionary processes of water-rich terrestrial planetary atmospheres, touching upon end-member states that are characterized either by uncontrolled sea ice albedo or by water vapor greenhouse feedbacks.

The sea ice albedo feedback can lead to rapid cooling whereupon the oceans become completely frozen over. Conversely, the water-vapor greenhouse feedback can lead to abrupt warming, water-rich atmospheres, and the total of loss of the oceans due to hydrodynamic escape or a thermal runaway. For many years, the leading descriptions of long timescale climatological evolution and the habitable zone have originated from energy balance and one-dimensional radiative-convective models (Budyko 1969; Hart 1979; Kasting et al. 1993; Selsis et al. 2007; Kopparapu et al. 2013). These works have shaped our thinking regarding the evolution of planetary climates. However, these models miss important feedbacks within the climate system caused by atmospheric dynamics, sea ice, clouds, and relative humidity.

Only recently have three-dimensional (3-D) climate system models become commonly used to place limits on the habitable zone (Abe et al. 2011; Boschi et al. 2013; Leconte et al. 2013; Shields et al. 2013, 2014, 2016; Yang et al. 2013, 2014; Wolf and Toon, 2014, 2015; Godolt et al. 2015; Kopparapu et al. 2016; Popp et al. 2016). In both 3-D and lower-dimensional models, the first objective has been to study an Earth-like planet, as Earth has the only climate system that is well observed and Earth is the only confirmed habitable world. As a first step, modern 3-D climate systems models have been applied to study an Earth-like exoplanet as it evolves across the habitable zone due to changing stellar luminosity. 3-D models allow for a self-consistent treatment of water in the climate system, including water vapor, clouds, surface ice, and oceans, and their respective spatial and temporal distributions about the planet. The presence or absence of each phase of water significantly affects the radiative energy budget of the planet, and thus the climate. However, there remains uncertainty amongst different 3-D Earth climate system models with regard to clouds (Flato et al. 2013), convection (Del Genio et al. 2016) and radiative transfer (Yang et al. 2016). For instance, amongst leading climate models, the increase in global mean surface temperature of the Earth in response to a doubling of $CO_2$ varies between 2.1 and 4.7 K (Andrews et al. 2012). Differences can become more significant for exoplanetary problems where the implied forcings tend to be larger (e.g. see Fig. 7a in Popp et al., 2016). We still have much to learn both scientifically and technically, as we apply our 3-D models to the new and exotic atmospheres of extrasolar planets.

Here we present simulations from a state-of-the-art 3-D climate system model of an Earth-like planet with a fixed amount of $CO_2$ in its atmosphere around F-, G-, and early K-dwarf main sequence stars, over a wide range of stellar fluxes. We do not consider M-dwarf star systems in this study. Habitable zone planets around these low-mass stars are likely to be tidally locked, which has a profound impact on planetary climate (Yang et al. 2013; Yang et al. 2014; Leconte et al. 2015; Kopparapu et al. 2016; Way et al. 2016). Planets in the habitable zone of F-, G-, and early K-dwarf stars, as studied here, fall outside the tidal locking radius and thus can rotate rapidly, like we observe for Earth and Mars. Venus, however, falls outside the tidal locking radius but is a slow rotator due to atmospheric tides.

Here, we determine the climate of rapidly rotating terrestrial planets under varying stellar fluxes, and provide constraints on the habitable zone under fixed $CO_2$ conditions. The interaction of atmospheric circulation, water vapor, clouds, and surface ice all play critical roles in modulating climate, and controlling sharp positive feedbacks. While earlier studies have similarly mapped climate as a function of stellar insolation using energy balance or 1-D models, to our knowledge this is the first such study to use an advanced, 3-D climate system model to attempt to map the entire range of habitable climates, complete from snowball to moist greenhouse, around numerous types of main sequence stars.

## 2. Methods

Here we use the Community Atmosphere Model version 4 (CAM4) from the National Center for Atmospheric Research (*Neale et al*. 2010). We build upon the prior work of Shields et al. (2013; 2014) and Wolf & Toon (2015) with new and complimentary simulations, facilitating a comprehensive description of the evolution of Earth-like climate through the habitable zones of F-, G-, and K-dwarf stars. Simulations of warm climates (*i.e.* those approaching a moist greenhouse) and cold climates (*i.e.* those approaching a snowball glaciation) follow the specific modeling methods described in detail in Wolf and Toon (2015) and Shields et al. (2013, 2014) respectively. Different setups for warm and cold simulations sets are used in order to combine new simulations with prior simulations of Wolf and Toon (2015) and Shields et al. (2013, 2014), saving considerable computational expense. In total, we use data from ~45 previous simulations from Shields et al. (2013; 2014) and Wolf & Toon (2015), and ~45 new simulations, yielding a complete picture of habitable zone climates.

There are some differences in the specific configuration of CAM4 used for warm and cold simulations sets, including resolution, ocean heat transport, land area assumptions, and the radiative transfer module used in the calculation (see Table 1). Of importance, cold simulations assume zero ocean heat transport and a global ocean, which allows for a transition to snowball Earth to occur at higher stellar fluxes than if some ocean heat transport is included (Poulsen et al. 2001; Pierrehumbert et al. 2011). Simulations approaching a moist greenhouse assume present day Earth continents and present day ocean heat transport. Cold simulations use the native CAM radiative transfer scheme found in CAM versions 4 and earlier (Ramanthan & Downey, 1986; Briegleb, 1992), while warm simulations use a correlated-*k* radiative transfer scheme (Wolf & Toon, 2013). Both configurations use identical atmosphere, ocean, and sea ice physics (Neale et al. 2010). However, despite these noted differences, results from each configuration are well in agreement where simulations overlap, for conditions near present day Earth surface temperatures (see Sections 3.1 and Fig. 2).

All simulations assume an Earth-like planet, with Earth's mass and radius, a 50-meter deep mixed-layer thermodynamic ("slab") ocean, and a 1-bar $N_2$ atmosphere with present day $CO_2$ concentrations but no ozone. Prognostic bulk microphysical parameterizations for condensation, precipitation, and evaporation control atmospheric water vapor, liquid cloud, and ice cloud condensate fields (Rasch and Kristjánsson, 1998). Deep convection (i.e., moist penetrative) is treated using the parameterization of Zhang and McFarlane (1995). Shallow convective overturning is treated by the parameterization of Hack (1994). We use stellar spectra from F2V (σ Bootis HD 128167), G2V (the Sun,), and K2V (ε Eridani HD 22049 and synthetic) stars (Segura et al. 2003; Pickles, 1998). In Fig. 1, we show F-, G-, and K- dwarf spectra normalized to 1360 W $m^{-2}$. These stars have effective temperatures ($T_{eff}$) of 6954 K, 5778 K, and 5084 K. In general F-, G-, and K-dwarf stellar classifications span $6{,}000 \leq T_{eff} \leq 7500$ K, $5200 \leq T_{eff} \leq 6000$ K, and $3{,}700 \leq T_{eff} \leq 5200$ K respectively (Habets & Heintze, 1981). We do not consider photochemistry here, however the stellar type, stellar activity, and atmospheric oxygen concentration are all critical for determining the UV radiation hazard for the planet surface, and have been studied elsewhere (Segura et al. 2003; Rugheimer et al. 2015). For each planet, we assume Earth-like orbital characteristics, with an orbital period of 1 Earth year, zero eccentricity, and 23.5° obliquity. Differences in the semi-major axis and orbital period that arise due to the respective mass and luminosity of each star are not incorporated. However, changes to the orbital periods do not appreciably affect global mean climate for rapidly rotating planets within the habitable zones of F-,

G- and early K-dwarf stars (Godolt et al. 2015). For this study we assumed a 24-hour rotation rate for all simulations. This is reasonable, as habitable zones studied here are outside the tidal spin-down region of their respective host stars (Leconte et al. 2015).

## 3. Results

### *3.1 Control Simulations*

First, we compare a set of standard atmospheres around F-, G-, and K-dwarf stars, using the model default present day Earth solar insolation of 1361.27 W $m^{-2}$ (Table **2**). In these simulations we assume the warm model configuration (Table 1), with a continental configuration and implied ocean heat transport that matches the present day Earth. The spectra emitted from cooler stars is relatively redder (Fig. 1), and thus interacts more strongly with atmospheric clouds, $CO_2$ and water vapor, as well as surface water ice and snow. Surface water ice (Dunkle & Bevans, 1956), as well as atmospheric clouds, $CO_2$ and water vapor (Kasting et al. 1993, Selsis et al. 2007) strongly absorb in the near-infrared. The end result is that the spectrally averaged all-sky planetary albedo is lower for an Earth-like planet around cooler stars (Table **2**). Thus it takes less stellar flux to warm an Earth-like planet around cooler stars. This albedo change is well established from simpler models and our results concur (Kasting et al. 1993; Kopparapu et al, 2013; Yang et al. 2014). The clear-sky (*i.e.* without clouds) albedo is reduced by a factor of ~2 between F- and K-dwarf control cases, owing to decreased Rayleigh scattering and increased overlap of the incident stellar spectra with near-infrared water vapor absorption bands. Meanwhile, the change in cloud albedo is relatively small across each different stellar type, varying by only about ~6% (analogously a cloud albedo change of ~0.01).

Nonetheless, in each control simulation the overall climate is not radically changed due to altering only the stellar spectra but not the stellar energy input (Table **2**). Climate does not switch states, remaining (generally) like the present-day Earth, dominated by open ocean but with some sea ice at the poles. The strength of the greenhouse effect and the global mean cloud fractions remain fairly similar for each control case. Note the greenhouse effect is given in units of temperature in order to facilitate comparison with Godolt et al. (2015), and is calculated using the Stefan-Boltzmann law from the difference in upwelling longwave radiation between the surface and the top-of-the-atmosphere implied by clouds and absorbing gases respectively. Differences in water vapor, cloud water, and sea-ice fractions come as no surprise, as these quantities are strongly dependent on the planetary temperature. Differences in the evolution of sea-ice, convection, clouds, and the distribution of relative humidity are significant drivers of planetary climate, and serve to amplify initial radiative perturbations due to albedo changes that are implied by changing the spectral energy distribution. For the F-dwarf star, cooling caused initially by increased atmospheric scattering and surface reflectivity is then amplified by the sea-ice albedo feedback. For the K-dwarf star, warming caused initially by increased absorption by water vapor and the surface is then amplified by the water vapor greenhouse, and is also linked to cloud feedbacks.

Interestingly, our results shown in Table 2, exhibit remarkably less sensitivity to changes in the stellar spectra compared to a similar set of simulations conducted with the $5^{th}$ version of the European Center Hamburg GCM (ECHAM5) from Godolt et al. (2015). Irradiated by the same F- and K-dwarf spectra, global mean surface 2-meter air temperatures reach 280.1 K and 294.1 K in CAM4, while in ECHAM5 they reach 273.6 K and 334.9 K respectively. Note that Popp et al. (2016) similarly find that ECHAM is significantly more sensitive than CAM4 to increasing stellar flux from the Sun. The striking differences in climate between these simulations highlights the need for detailed model intercomparison to ascertain why the models diverge. Without further work, we cannot determine if the model differences arise purely from differences in radiative transfer, or whether sea-ice and cloud feedbacks are more to blame.

For simulations near the present day surface temperatures, model configurations for warm and cold simulations (see Table 1) yield very similar results, with mean surface temperatures generally within ~2 K for temperate conditions (see Fig. 2). Under the present day stellar flux from the Sun, a G-dwarf star, the cold (warm) configurations yield a global mean surface temperature of 287.3 K (289.0 K). Under insolation from an F-dwarf star, a 5% increase in the solar constant above the present day is required to reach approximately modern day surface temperatures of 288.4 K (287.1 K) for the cold (warm) configuration. For the K-dwarf star case, a 2% reduction in the solar constant yields 290.3 K (291.6 K) for the cold (warm) configurations. However, at colder temperatures differences emerge between the warm and cold configurations due primarily to ocean heat transport , which is assumed not to occur in the cold configuration. At the present day solar flux, F-dwarf simulations become cold (241.6 K) when no ocean heat transport is included. This behavior in response to turning off ocean heat transport is in agreement with Godolt et al. (2015).

### *3.2 Multiple climate states*

Fig. 2 shows the evolution of global mean surface temperature ($T_s$) and climate sensitivity for an Earth-like planet with fixed $CO_2$ as a function of the relative stellar flux from F-, G-, and K-dwarf stars. Relative stellar flux is defined as the ratio between the incident stellar flux on the planet, and that received by Earth at present day ($S_0$), taken here to be 1360 W m$^{-2}$. Thus in Fig. 2 a relative stellar flux ($S/S_0$) of 1.0 equals an incident stellar flux on the planet of 1360 W m$^{-2}$, approximately matching the present day Earth value of 1361 W m$^{-2}$. Four stable climatic regimes are indicated by shaded regions in Fig. 2; snowball ($T_s \leq 235$ K), waterbelt (235 K $\leq T_s \leq$ 250 K), temperate (275 K $\leq T_s \leq$ 315 K) and moist greenhouse ($T_s \geq 330$ K). Stable climates are in equilibrium, having balanced incoming and outgoing radiation and have no systematic temperature drift. The evolution of climate around each type of star is qualitatively similar. Note that here we consider the moist greenhouse climate state to be defined by the radiative-convective state of the atmosphere, as described by Wolf and Toon (2015). Wolf & Toon (2015) found that the climate undergoes an abrupt transition between temperate and moist greenhouse states, characterized by the closing of the 8 to 12 μm water vapor window region, increased solar absorption in the near-infrared water vapor bands, and the subsequent stabilization of the low atmosphere against convection. We discuss water-loss rates versus the depletion of the ocean inventory separately below. Under a relaxed constraint, where any amount of surface water may constitute a habitable planet, waterbelt, temperate and cooler moist greenhouse states are habitable. However, in our surveys of the stars we seek not just to find habitable zone planets, but to find those that are preferably within the temperate climate regime, where the vast majority of the planet is ice-free and temperatures are similar to the Earth presently.

The four stable climate regimes are separated by sharp climatic transitions, indicated by maxima in climate sensitivity (numbered in Fig. 2d, e, f). Climate sensitivity is the change in global mean surface temperature for a given change in radiative forcing, here from incrementally increasing or decreasing the incident stellar flux. The sharp climatic transitions are triggered by interactions between sea-ice, water vapor, as well as clouds and radiation, whereupon a small change in solar forcing can beget a large change in $T_s$. Interestingly, for some temperature ranges, 250 K $\leq T_s \leq$ 275 K and 315 K $\leq T_s \leq$ 330 K, stable climate states never occur in our model. These unallowable temperature regions are caused by uncontrolled sea-ice albedo, and uncontrolled water-vapor greenhouse feedbacks respectively. Each cause climate sensitivity to spike (Fig. 2). If 250 K $\leq T_s \leq$ 275 K, climatic stability can be re-established either by warming into the temperate state where sea-ice is trapped at the poles, or by cooling into the waterbelt state which is stabilized by albedo contrasts between bare sea ice and snow covered areas formed when global ice sheets encroach into the subtropical desert zone (e.g. Abbot et al. 2011). If 315 K $\leq T_s \leq$ 330 K, climate stability can be re-established either by cooling into the temperate state where the planet's surface can efficiently cool to space through the water vapor window region, or by warming into the moist greenhouse state which is stabilized by a reduction to relative humidity and the formation of an upper atmosphere cloud deck (Wolf & Toon, 2015). While these unallowable regions appear similarly for an Earth-like planet around each star, it is unclear how sensitive these regions may be to other choices of model parameters.

Earth presently exists in a region where two climate states are possible. In Earth's present temperate state (perhaps fortuitously) climate sensitivity is near a minimum against both positive and negative radiative forcings (Fig. 2e). The relative long-term stability of Earth's climate may be circumstantial evidence that terrestrial climates preferentially relax towards climate sensitivity minima. As shown in Fig. 2 and Shields et al. (2014), there is strong hysteresis between the snowball and temperate climate states, which has long been recognized from simple climate models (Budyko, 1967). The solar constant must be raised to high levels to escape from a snowball, but at much lower solar constants temperate climate states are stable, but only if the climate was initially warm. Thus there is a range of solar fluxes at which climate exhibits bistability, with both snowball or temperate and waterbelt states being possibly stable at given stellar fluxes. The actual state depends on the initial conditions, and thus the planet's evolutionary history. For planets around F-, G-, and K-dwarf stars, Earth-like planets exhibit bistability for relative stellar fluxes ($S/S_0$) of 0.99 – 1.14, 0.92 – 1.06, and 0.88 – 0.98 respectively. Note that the extent of the bistable region is encouragingly quite similar to that found from 3-D models of intermediate complexity (Lucarini et al. 2010; Boschi et al. 2013), which found a bistable range of 0.93 – 1.05 for Earth around the Sun. Bistability is not found in our calculations between temperate and the moist greenhouse climate states. However, Popp et al. (2016) find evidence of a small (*i.e.* contained in a ~2% change in stellar flux) hysteresis between temperate and moist greenhouse climates using an idealized version of the ECHAM6 climate model.

The runaway greenhouse provides the most generous bounds for the inner edge of the habitable zone for an Earth-like planet. Kasting et al. (1993) also defined the more conservative "moist greenhouse" inner edge of the habitable zone based on water-loss to space from moist atmospheres, which may occur at lower stellar fluxes than are needed to induce a thermal runaway. Kasting et al. (1993) define this inner edge constraint to occur where diffusion limited escape causes the entirety of Earth's oceans ($1.4 \times 10^{24}$ g $H_2O$) to be lost to space in a period of time that approaches the present age of the Earth. In practical terms, this threshold is reached when the stratospheric $H_2O$ volume mixing ratio equals $3 \times 10^{-3}$. Here we adopt a marginally more stringent constraint, assuming that an Earth ocean of water must be lost within ~1 Gyr. This ensures that the water-loss timescale is meaningful even for F-dwarf stars, which live little more than half as long as our Sun (Rushby et al. 2013). This escape rate occurs when the stratospheric water vapor volume mixing ratio exceeds $\sim 7 \times 10^{-3}$, which occurs in our model when $T_s \sim 355$ K around each star (Fig. 2). However, note that Earth-like planets with $T_s \sim 350$ K will desiccate within about ~8 Gyr, which is near the main sequence lifetime of G-dwarf stars and significantly less than that of K-dwarf stars.

Fig. 3 shows the model top (~0.2 mb) temperature and water vapor mixing ratio as a function of mean surface temperature, as climate warms around each star. The temperature controls the amount of water vapor in the upper atmosphere. At low $T_s$ the upper atmosphere is noticeably warmer around redder stars due to the effect of increased absorption by water vapor in the near-infrared, coupled with inefficient radiative cooling aloft. However, this trend becomes muted for increasing $T_s$ as the atmospheres become increasingly water-rich, thermally opaque, and convective to high-altitudes (Wolf & Toon, 2015). The time-scale to lose Earth's oceans falls off dramatically as the mean surface temperature increases. By the time $T_s \sim 360$ K, the oceans would be lost in only several hundred million years. Such atmospheres would be short-lived relative to stellar lifetimes, and would transition into dry planets (Abe et al. 2011). However, relatively cooler moist greenhouse atmospheres, with 330 K $\leq T_s \leq$ 350, have upper atmosphere water vapor volume mixing ratios between ~$10^{-6}$ and $10^{-3}$, and thus can retain an Earth ocean of water for tens to hundreds of billion of years. More detailed hydrodynamic escape calculations from these atmospheres would better our understanding of habitable lifetimes of planets near the inner edge of the habitable zone. Furthermore, on other planets hydrogen escape rates may vary due to factors we have not considered such as differences in the exobase temperature, the mean molecular weight of the atmosphere, the gravitational force, stellar activity, and photochemistry.

### *3.3 Circumstellar Climate Zones*

Based on the climate results shown in Fig. 2, we can define circumstellar climate zones for Earth-like planets at modern-day $CO_2$ levels around F-, G-, and K-dwarf stars (Fig. 4). Here, circumstellar climate zones provide a more detailed description of habitable planetary climates than does the habitable zone. The habitable zone is based on the existence of liquid water, but does not take into account such issues as the ability of organisms to survive at a given temperature, or the climatological history of the planet. Significant differences in climate zones exist depending upon whether one assumes warm (*i.e.* ocean covered, Fig. 4a) or cold (*i.e.* completely ice covered, Fig. 4b) initial conditions. Following the practice of some recent habitable zone studies (Selsis et al. 2007; Zsom et al. 2013; Kopparapu et al. 2013, 2014; Yang et al. 2014), based on our climate modeling results we determine parametric relationships between the relative incident stellar flux received by an Earth-like planet at modern $CO_2$ that yields a given climate state ($S_{climate}$) noted in Fig. 4, and the stellar effective temperature ($T_{eff}$) of the host star. Equations 1 and 2 are valid for stars with $T_{eff}$ between 4900 K and 6600 K,

1) $S_{climate} = a + bT_* + cT_*^2$

2) $T_* = T_{eff} - 5780$ K

where coefficients *a*, *b*, and *c,* are given in Table 3 for each particular climate. The results of $S_{climate}$ using Table 3 describe circumclimate zones that are plotted and labeled in Fig. 4.

All circumstellar climate zones found in Fig. 2, are shown in Fig. 4 as shaded regions bounded by solid lines. Additional climate states of interest are indicated by dashed lines. First, moving from left to right in Fig. 4a (from high $T_s$ and high solar flux), we delineate with a dashed line the inner edge of the habitable zone according to the ~1 Gyr water-loss criteria ($T_s$ ~ 355 K) discussed in section 3.1. Next we mark with a solid orange line the radiative-convective transition from a moist greenhouse climate state into a temperate state, as described by Wolf and Toon (2015), whereupon $T_s$ abruptly drops from ~330 K to ~315 K. Next we mark the heat stress limit for mammalian biological functioning. Sherwood and Huber (2010) argue that if $T_s \geq 300$ K, then the majority of the Earth's human population would be subject to prolonged periods of lethal heat stress. While technological and biological adaption may facilitate survival at these hotter temperatures, life and society as we know them would be threatened. Next, we mark the climatic conditions of the present day Earth ($T_s$ ~289 K) around each star. In Fig. 4a we define with a solid dark blue line, the boundary where an Earth-analog planet with identical $CO_2$ would transition from the temperate regime into a waterbelt (*i.e.* ice lines reaching the tropics), a condition only accessible via cooling from a warmer state. Finally, we define with a solid light blue line the transition into a snowball Earth state via reduced solar forcing. If formed initially warm, a planet may access a temperate zone with $\Delta S/S_0 \sim 0.16 - 0.20$ (equivalently a ~218 – 272 W $m^{-2}$ change in total solar insolation received by the planet) as marked by the green shaded region in Fig. 4a. Here, $\Delta S/S_0$ is the width of the climate zone in units of relative stellar flux and varies for planets around different effective temperature stars. The habitable zone, including the waterbelt state (for warm start only) up to the water-loss threshold, spans $\Delta S/S_0 \sim 0.24 - 0.34$ (~326 – 462 W $m^{-2}$) for initially warm planets.

In the absence of $CO_2$ changes, if a planet is initially cold (i.e. fully ice covered), its habitable zone is constrained by the solar deglaciation and water-loss limits indicated in Fig. 4b, and there is a relatively small range of possible habitable states. Note that a present day Earth-like climate cannot be accessed from a cold start (Fig. 4b, see also Fig. 2). Moving from right to left (from low $T_s$ and low solar flux), solar driven deglaciation is indicated by a light blue line in Fig. 4b. The temperate climate zone is significantly narrower with 290 K $\leq T_s \leq$ 315 K, and the waterbelt state is skipped entirely from a cold start. There is no difference between the inner edge of the habitable zone for cold and warm start cases. By the time that climate has warmed to moist greenhouse and water-loss thresholds, snow and ice have long since vanished from the planet, and no memory of the cold initial conditions remain. Thus, the temperate climate regime spans only $\Delta S/S_0 \sim 0.06 - 0.11$ (~82 – 150 W $m^{-2}$) as marked by the green shaded region in Fig. 4b. The full width of the habitable zone for initially cold planets is marked from solar deglaciation to water-loss limits, and spans $\Delta S/S_0 \sim 0.14 - 0.19$ (~190 – 258 W $m^{-2}$). It is interesting to note that the habitable zone is wider for a cold initial planet around a K-dwarf star. This is because solar driven deglaciation is more effective around relatively redder stars, due to the low near-infrared albedo of snow and ice (Joshi & Haberle 2012; Shields et al. 2014).

### 3.4 Habitable lifetimes

Another consideration in Fig. 2 and Fig. 4 is the time that is spent in the habitable zone. Main sequence stars brighten over the course of their lifetime, thus the radiation received by an orbiting planet increases over time (Iben 1967). However, the main sequence lifetime and rate of luminosity evolution depends upon the stellar type. If we assume that the rate of luminosity evolution is linear in time, we can then make simple estimates for the lifetime of climate zones ($\tau_{climate}$), calculated as the time needed for the stellar luminosity to evolve corresponding to the maximum width of each climate zone in terms of the relative stellar flux ($\Delta S/S_0$) shown in Fig. 4. We use equation 3:

$$3)\ \tau_{climate} = \frac{\Delta S/S_0}{\frac{L_{tms}}{L_0} - \frac{L_{zams}}{L_0}} \times \tau_{ms}$$

where $L_{tms}/L_0$ is the luminosity at the end of the star's main sequence lifetime normalized to the present solar luminosity, $L_{zams}/L_0$ is the zero age main sequence luminosity normalized to the present solar luminosity, and $\tau_{ms}$ is the total main sequence lifetime of the star. $L_{tms}/L_0$ and $L_{zams}/L_0$ are calculated using the parametric fits given by equations 3 and 5 in Guo et al. (2009). Following Rushby et al. (2013, equations 7 and 8), the total main sequence lifetime can be computed as:

$$4)\ \tau_{ms} = 10.9 * \frac{M_0}{M^3}$$

where $M$ is the mass of the star, $M_0$ is the mass of the Sun, and 10.9 is the main sequence lifetime of our Sun in billions of years. Our calculation of $\tau_{climate}$ assumes that the planet initially has the lowest temperature allowed for a given climate zone, and is then warmed via main sequence brightening through the climate zone. Thus $\tau_{climate}$ is the maximum length of time that an Earth-like planet, at fixed $CO_2$, could remain within a given climate zone under the influence of the main sequence brightening. Beginning its life at a higher temperature would decrease $\tau_{climate}$, while a draw down of $CO_2$ could possibly lengthen $\tau_{climate}$.

In Fig. 5 we consider $\tau_{climate}$ for the temperate climate zone, and for the habitable zone in total, including waterbelt and moist greenhouse states with $T_s$ below 355 K. Viewed in this fashion it is clear that lower effective temperature stars (*i.e.* the K-dwarf) provide a more stable climatic environment because $\tau_{climate}$ for K-dwarfs is about double that of the F- and G-dwarf stars. While habitable planets around K-dwarf stars are more sensitive to changes in the stellar flux than planets around F- and G-dwarfs (Fig. 2a,d), their long main sequence lifetimes (~19.8 Gyr for ε Eridani for instance) and thus slower temporal luminosity evolution across the main sequence bestows a significant advantage for evolution of life. The most optimal scenario is for initially warm planets. Terrestrial planets are believed to have been formed hot from accretion, and with an initially molten surface before their earliest atmospheres cooled and condensed. Thus, even though stellar luminosity increases in time, terrestrial planets likely begin their earliest histories in a hot state, and could then access a wider temperate zone, and undergo waterbelt states upon first cooling (Fig. 4). Still, a waterbelt state may be a low probability occurrence due to the narrow range of allowable stellar fluxes. Alternatively, waterbelt states could also occur if a planet formed with a larger primordial $CO_2$ inventory than assumed here, which is then drawn down over time by weathering processes, allowing the planet to cool (Abbot et al. 2011).

The maximum time spent in the habitable zone is ~2.2, ~2.4, and ~4.7 Gyr for F-, G-, and K-dwarf planets respectively, possible only for warm start scenarios. Note that life has existed on Earth for at least 3.8 gyr (Nisbet & Sleep, 2001), significantly longer than the 2.4 Gyr lifetime noted here for a G-dwarf star. Long-lived habitable conditions for Earth are most likely due to a stronger $CO_2$ greenhouse early in Earth's history. See section 4 for more discussion. Cold start cases apply to initially frozen worlds subject to increased stellar fluxes whether by stellar evolution, or possibly planetary migration. For cold initial conditions, temperate climate states around F- and G-dwarf stars may last only ~500 Myr, if a draw down of $CO_2$ is not invoked to mitigate warming. Interestingly, the habitable and temperate climate zone lifetimes are only about ~10 − 20% shorter for F-dwarf planets compared to the G-dwarf, despite the F-dwarf having a total main sequence lifetime that is ~40% shorter (6.4 Gyr versus 10.9 Gyr). Temperate Earth-like planets around F-dwarf stars benefit from their bluer stellar spectra, which is more effectively scattered and less readily absorbed by the near-IR water vapor bands, thus allowing for a temperate climate zone that exists under a wider range of relative stellar fluxes (Fig. 2 and Fig. 4). Thus we should not ignore habitable zone planets around F-dwarf stars, due to their muted climate sensitivity.

**4. Discussion**

In this study we have fixed $CO_2$ at present day values, thus Fig. 2 and Fig. 4 illustrate a relative range of habitable climates for a single choice for non-condensable greenhouse species. Couched in these terms, the habitable zone appears quite narrow for both warm and cold start scenarios. The wide range of stellar fluxes possible for the conventional habitable zone, as described in Kopparapu et al. (2013), is reliant on the strong greenhouse effect from multi-bar $CO_2$ atmospheres to set the outer edge of the habitable zone. It is conventionally thought that $CO_2$ can change on many planets and provide a stabilizing feedback to the climate system through the temperature dependent action of the carbon/silicate cycle. If the temperature increases for some reason, weathering rates will increase, leading to removal of $CO_2$ from the atmosphere/ocean system and sequestration into sea floor carbonates such as limestone, which cools the planet by weakening the total greenhouse effect. Subduction of carbonate-rich sea floor on planets with plate tectonics and subsequent volcanic outgassing recycles the sea floor carbonates and supplies $CO_2$ to the atmosphere to balance the loss from weathering over geologically long periods of time. Of course, silicate weathering requires the presence of continents and plate tectonics, both of which are uncertain.

However, if a planet has weak volcanic outgassing, perhaps in tandem with high rates of formation of seafloor carbonates, a cold planet may not be able accumulate sufficient $CO_2$ to deglaciate the planet independently of the stellar flux. Mars, for instance, currently resides within the habitable zone of Kopparapu et al. (2013), but it was unable to retain sufficient $CO_2$, to escape from its present frozen state. Kadoya and Tajika (2015) and Haqq-Misra et al. (2016) argue for Earth that if the paleo-$CO_2$ outgassing rates were less than on Earth presently, carbon/silicate cycles may not have been able to prevent a snowball glaciation for most of Earth's history. Furthermore, planets entering a snowball phase may oscillate between frozen and thawed states, with a frequency dependent on the rate of outgassing (Tajika, 2007; Mills et al., 2011; Driscoll & Bercovici, 2013; Haqq-Misra et al., 2016). Note while the early Earth was indeed habitable for nearly its entire existence, implying higher $CO_2$ outgassing rates in the distant past to sustain a warm climate, there is geological evidence for periodic snowball glaciations up through the Neoproterozoic period, with the last being ~635 Myr ago, when the solar constant was at ~94% of its present day value (Kirschvink et al. 2000; Pierrehumbert et al. 2011). Interestingly, while simple life emerged billions of years earlier, complex life did not emerge until the quasi-periodic snowball events ceased (Hoffman et al., 1998). While here we emphasize $CO_2$ and carbon/silicate cycles, there could be numerous other gases and particles in planetary atmospheres that could impact their climate that we have not considered, including $N_2$, $H_2$, $CH_4$, organic hazes, various sulfur compounds, and a host of others.

In Fig. 6 we summarize habitable zone calculations published to date using 3-D climate models of Earth-like planets, along with the widely used values of Kopparapu et al. (2013; 2014) that are based on 1-D radiative-convective model calculations. Fig. 6a shows constraints on the inner edge of the habitable zone derived from climate models with solid lines and diamonds. In theory, the inner edge of the habitable zone for a water-rich planet should not involve high levels of $CO_2$. As the climate warms, enhanced silicate weathering should draw down $CO_2$ to relatively low levels. Each simulation shown assumes a $CO_2$ mixing ratio equal to the modern Earth, along with 1 bar $N_2$ as the background gas. Numerous studies have calculated the inner edge of the habitable zone for Earth around the Sun ($T_{eff}$ = 5778 K) using nationally supported 3-D climate system models (Abe et al. 2011**;** Leconte et al. 2013; Wolf &Toon, 2014, 2015; Yang et al. 2014; Popp et al. 2016). The models vary in the location of the inner edge of the habitable zone by ~15% ($S/S_0$). The models also differ in the predicted end state of the atmosphere. Leconte et al. (2013), using the LMD generic climate model, predict Earth will enter a runaway greenhouse and no moist greenhouse state is possible. On the contrary both CAM (Wolf & Toon, 2015) and ECHAM (Popp et al. 2016) models predict that a climatologically stable moist greenhouse state with significant water-loss marks the inner edge of the habitable zone for Earth. Note that the results from this study using a modified version of CAM4 appear quite similar to those of Yang et al. (2014), who used CAM3 (marked "fast" on Fig. 6). However, the similarity between the two results is somewhat deceiving. Yang et al. (2014) derived the inner edge of the habitable zone at a point where the model becomes numerically unstable ($T_s$ ~ 310 K). Here numerical improvements to the convection scheme allow us to simulate much hotter temperatures, and define a true inner edge of the habitable zone by water-loss from moist greenhouse atmospheres. Note also that the position of the inner edge determined by Yang et al. (2014) and Wolf & Toon (2015) is also affected by differing properties of the radiative transfer model used (see Yang et al. 2016).

In all panels of Fig. 6 light yellow and light blue shaded regions indicate regions in space that are inside and outside of the tidal locking radius respectively, following Edson et al. (2011) and Haqq-Misra & Kopparapu (2015). Tidal locking is expected to be most important for planets towards the inner edge of the habitable zone around low mass stars. Planets located within the tidal locking radius are strongly influenced by the host star gravity, and should exist in synchronous or resonant orbital-rotational configurations. Although, tidal spin down may also be dependent upon atmospheric thickness and thermal tides (Leconte et al. 2015). Planets located in the blue region are unconstrained and can rotate rapidly as does Earth and Mars. There is a large difference between constraints for the inner edge of the habitable zone for rapidly rotating planets (Yang et al. 2014; Leconte et al. 2013; Godolt et al. 2015; Wolf and Toon, 2015) and tidally locked planets, which tend to be more slowly rotating (Yang et al. 2014; Kopparapu et al. 2016). As first described by Yang et al. (2013), strong convection at the substellar point of slow rotators can create thick clouds that substantially raise the planet's albedo, thus allowing liquid surface water to exist under large stellar fluxes. Shown here are calculations from Yang et al. (2014) using CAM3 which assumed a fixed orbital-rotational period of 60 Earth days (marked "slow" on Fig. 6a). However, in reality the orbital-rotation period and the stellar flux received by a tidally locked planet are dependent on the mass and luminosity of the stars, and should vary between ~10 and 65 days for planets near the inner edge of the habitable zone around late-K and early M stars. A subsequent revision of the inner edge of the habitable zone for these slow rotators was published by Kopparapu et al. (2016) using CAM4 and self-consistent orbital periods for both high and low metallicity stars. Changing the planetary rotation rate self-consistently has important consequences for atmospheric dynamics, cloud fields, and ultimately the global mean albedo and climate. Seamlessly connecting the habitable zones for rapid and slow rotators is not trivial. Near $T_{eff} \sim 4500$ K, the inner edge of the habitable zone for rapid rotators approaches the tidal locking radius. Here one may find a transition region, between the fast and slow rotator limits, dependent upon planet-star tidal interactions and the precise rotation rate of the planet in question. It is clear from Fig. 6a that the evolution of climate may critically depend on the evolution of the planetary rotation rate due to tidal interactions with the host star and also moons.

In Fig. 6b we summarize constraints for the outer edge of the habitable zone and for habitable climates at low stellar fluxes and high-$CO_2$, with dashed lines and triangles. We include our estimates for the outer edge of the habitable zone based on snowball glaciation and solar driven deglaciation limits for an Earth-like planet with modern $CO_2$. Note in Fig. 6b, the "snowball" limit marks the transition between habitable and globally ice covered states due to decreasing solar insolation (see also Fig. 4a). The "deglaciation" limit marks the transition between globally ice covered and habitable states triggered by increasing solar insolation (see also Fig. 4b). Kasting et al. (1993) first postulated the so-called maximum $CO_2$ greenhouse limit for the outer edge of the habitable zone using a 1D radiative convective model. This limit has recently been revised for initially warm (Kopparapu et al. 2013), and initially ice covered planets (Haqq-Misra et al. 2016), using similar methodology with a 1D model. To date no 3-D simulations have calculated the maximum $CO_2$ greenhouse limit. However, several 3D studies have explored paleo-Earth, paleo-Mars, and high-$CO_2$ exoplanets scenarios, which may serve as useful steps towards determining the outer edge of the habitable zone around different stars (Wordsworth et al. 2011; Urata & Toon, 2013; Forget et al. 2013; Wordsworth et al. 2013; Wolf & Toon, 2014; Shields et al. 2016). Several of these data points are included on Fig. 6b along with the $CO_2$ burden required to yield a habitable climate. More work is needed in defining the maximum-$CO_2$ greenhouse for terrestrial planets using 3-D models. Finally, Fig. 6c combines habitability constraints from Fig 6a and 6b onto the same plot. It is clear that without the ability to regulate $CO_2$ or other greenhouse gases, the habitable zone for rapidly rotating planets is quite narrow. Still, the effect of slowing rotation as an equally large effect in widening the total habitable zone. Information regarding planetary rotation rate, $CO_2$ cycling, and the ability of a planet to retain its atmosphere against escape are equally as important for determining habitability, as is the incident stellar flux.

## 4. Conclusions

Here we have used a 3-D climate system model to explore the effect of changing stellar fluxes on climate for an Earth-like exoplanet around F-, G-, and early K-dwarf main sequence stars, assuming a fixed amount of $CO_2$. For these stars, the inner edge of the habitable zone lies beyond the tidal-locking radius, and thus planets are free to maintain a rapid rotation rate. Earth-like planets in the habitable zone are subject to four stable climate states (snowball, waterbelt, temperate, and moist greenhouse), each separated by sharp climatic transitions which are triggered by the changing thermodynamic partitioning of water in the climate system. Without allowing for the build up or removal of non-condensable greenhouse gas species such as $CO_2$, the range of relative stellar fluxes that permit temperate climates (*i.e.* $275 \leq T_s \leq 315$ K) is quite narrow; $\Delta S/S_0 \sim 0.06 - 0.11$ for initially frozen planets, and $\Delta S/S_0 \sim 0.16 - 0.20$ for initially warm planets. The range of habitable climates becomes marginally wider if we generously allow both waterbelt and cooler moist greenhouse climates ($T_s \leq 330$ K) to be included as habitable worlds; $\Delta S/S_0 \sim 0.14 - 0.19$ for initially cold planets, and $\Delta S/S_0 \sim 0.24 - 0.34$ for initially warm planets. For cold initial conditions, planets around K-dwarf stars have the widest habitable zones due to deglaciation of sea-ice at lower stellar fluxes. For warm initial conditions, planets around F-dwarf stars have the widest habitable zones due primarily to a muted climate sensitivity across a broad temperate climate zone. These variations in the solar constant represent only a small fraction of the change in solar constant over the host star's main sequence lifetime. Amongst our studied systems, the K-dwarf stars provides the longest lived habitable climates due to their lengthy main sequence lifetimes, and thus relatively slow rate of main sequence brightening.

The reader is also reminded that results presented in this work in Fig. 2, Fig. 4, and Fig. 6 are derived from a single three-dimensional climate system model. Differences in radiative transfer, convection, clouds, ocean heat transport, sea ice, and other processes can vary across different 3-D models, and can lead to significant differences in the resultant climates. Furthermore, we have only studied two parameters (stellar flux and spectrum) in detail. The computational expense of modern climate models requires a focused approach to the study of parameter spaces relevant to habitable extrasolar planets. Lower dimensional models retain significant value by allowing multi-dimensional parameter sweeps with relative ease. The standard approximation taken here of the proverbial Earth-like exoplanet is now well worn. We hope these simulations mark an appropriate starting point for intercomparison amongst current climate models for Earth-like planets around various stars, before continuing towards habitable planets having more exotic characteristics. Model intercomparison is needed to constrain the origin of differences found in various simulations some of which are noted in Fig. 6. The differences might arise from different model parameterizations of radiative transfer, clouds, convection, large-scale dynamics, or some other process.

While much effort in the literature has been given to defining the effect of the solar constant on habitable climates, it is clear that the geological and/or biological regulation of non-condensable greenhouse species is of equal or possibly greater importance to planetary habitability, by allowing the habitable zone to be extended much further away from the host star. Lastly, to complete our picture of the inner edge of the habitable zone, future work might focus on the 4500 K to 5000 K effective temperature regime, where the inner edge of the habitable zone for rapid rotators approaches the tidal-locking radius.

**Acknowledgements.** E.T.W. and O.B.T. acknowledge support from NASA Planetary Atmospheres Program award NNX14AH17G. A.S. acknowledges support from the National Science Foundation under Award No. 1401554, and from the University of California President's Postdoctoral Fellowship Program. R.K.K., E.T.W., and J.H. acknowledge funding from the NASA Habitable Worlds program under award NNX16AB61G. R.K.K. acknowledge funding from NASA Astrobiology Institute's Virtual Planetary Laboratory team, supported by NASA under cooperative agreement NNH05ZDA001C. This work utilized the Janus supercomputer which is supported by the National Science Foundation (award CNS-0821794) and the University of Colorado at Boulder. We would like to acknowledge high-performance computing support from Yellowstone (ark:/85065/d7wd3xhc) provided by NCAR's Computational and Information Systems Laboratory, sponsored by the National Science Foundation. This work was performed as part of the NASA Astrobiology Institute's Virtual Planetary Laboratory Lead Team, supported by the National Aeronautics and Space Administration through the NASA Astrobiology Institute under solicitation NNH12ZDA002C and Cooperative Agreement Number NNA13AA93A.

**References**

Abe, Y., Abe-Ouchi, A., Sleep, N. H., & Zahnle, K. J. 2011, AsBio, 11(5), 443

Abbot, D. S., Voigt, A., & Koll, D. 2011, JGR, 116, D18103

Andrews, T., Gregory, J.M., Webb, M.J., & Taylor, K.E. 2012, GRL, 39, L09712

Boschi, R., Lucarini, V., & Pascale, S. 2013, Icar, 226, 1724

Boyajian, T. S., McAlister, H. A., van Belle, G., et al. 2012, ApJ, 757, 112

Budyko, M. I. 1969, Tell, 21, 611

Butler, R. P., Wright, J. T., Marcy, G. W., et al. 2006, ApJ, 646, 505

Briegleb, B. P. 1992, JGR, 97, 7603

Del Genio, A. D. 2016, ArXiv e-prints [arXiv:1603.07424]

Driscoll, P. & Bercovici, D. 2013, Icar, 226, 1447

Dunkle, R. V. & Bevans, J. T. 1956, J. Meteorol., 13, 212

Edson, A., Lee, S., Bannon, P., et al. 2011, Icar, 212, 1

Flato, G., Marotzke, J., Abiodun, B., et al. 2013, in Climate Change 2013: The Physical Science Basis. Contribution of Working Group I to the Fifth Assessment Report of the Intergovernmental Panel on Climate Change, eds. T. Stocker, D. Qin, G.-K. Plattner, et al. (Cambridge, UK, New York, NY, USA: Cambridge University Press), 741

Forget, F., Wordsworth, R., Millour, E., et al. 2013, Icar, 222, 81

Gardner, J. P., Mather, J. C., Clampin, M., et al. 2006, SSRv, 123, 485

Godolt, M., Grenfell, J. L., Haman-Reinus, A., et al. 2015, P&SS, 111, 62

Godolt, M., Grenfell, J.L., Kitzmann, D., et al. 2016, A&A, 592, A36

Hack, J. J. 1994, JGR, 99, 5551

Habets, G. M. H. J. & Heintze, J. R. W. 1981, A&AS, 46, 193

Hansen, J., Sato, M., Ruedy, R., et al. 2005, JGR, 110, D18104

Hart, M. H. 1979, Icar, 37, 351

Haqq-Misra, J. & Kopparapu, R. K. 2015, MNRAS, 446, 428

Haqq-Misra, J., Kopparapu, R. K., Batalha, N. E., et al. 2016, ApJ, 827, 2

Hoffman, P. F., Kaufman, A. J., Kalverson, G. P., & Schrag, D. P. 1998, Sci, 281, 1342

Iben, I. 1967, ARA&A, 5, 571.

Joshi, M. M. & Haberle, R. M. 2012, AsBio, 12, 3

Kadoya S., & Tajika, E. 2015, ApJL 815, L7

Kasting, J. F., Pollack, J. B., & Ackerman, T. P. 1984, Icar, 57, 335

Kasting, J. F. 1988, Icar, 74, 462-494

Kasting, J. F., Whitmire, D. P. & Reynolds, R. T. 1993, Icar, 101, 108

Kirschvink, J. L., Gaidos, E. J, Bertani, L. E. et al. 2000, PNAS, 97, 1400

Kitzmann, D., Alibert, Y., Godolt, M., et al. 2015, MNRAS 452, 3752

Kopparapu, R. K., Ramirez, R., Kasting, J. F., et al. 2013, ApJ, 765, 131

Kopparapu, R. K, Ramirez, R. M., Kotte, J. S., et al. 2014, ApJL, 787, L29.

Kopparapu, R. K., Wolf, E. T., Haqq-Misra, J., et al. 2016, ApJ 819, 84

Leconte, J., Forget, F., Charnay, B. et al. 2013 Natur, 504, 268

Leconte, J., Wu, H., Menou, K., & Murray, M. 2015, Sci, 347,632

Lucarini, V., Fraedrich, K., & Lunkeit, F. 2010, QJRMS, 136(646), 2

Mills, B., Watson, A. J., Goldblatt, C., et al., 2011, Nat. Geosci., 4, 861

Neale, R. B., Richter, J. H., Conley, A. J., et al. 2010, NCAR/TN-486+STR

Nisbet, E. G. & Sleep, N. H. 2001, Natur, 409, 1083

Pickles, A. J. 1998, PASP, 110, 863

Pierrehumbert, R. T., Abbot, D. S., Voigt, A. & Koll, D. 2011, AREPS, 39, 417

Popp, M., Schmidt, H., & Marotzke, J. 2016, NatCo, 7, 10627

Poulsen, C. J., Pierrehumbert, R. T., & Jacob, R. L. 2001, GRL, 28, 1575

Ramanathan, V. & Downey, P. 1986, JGR, 91, 8649

Rasch, P. J. & Kristjánsson, J. E. 1998, J. Clim, 11, 1587

Ricker, G. R., Winn, J. N., Vanderspek, R., et al. 2014, Proc. SPIE, 9143, 914320

Rugheimer, S., Segura, A., Kaltenegger, L., & Sasselov, D. 2015, ApJ, 806, 137

Rushby, A. J., Claire, M. W., Osborn, H., & Watson, A. J. 2013, AsBio, 13(9), 833

Segura, A., Krelove, K. Kasting, J. F., et al. 2003, AsBio, 3, 689

Selsis, F., Kasting, J. F., Levrard, B., et al. 2007, A&A, 476, 1373

Sherwood, S. C. & Huber, M. 2010, PNAS 107(21), 9552

Shields, A. L., Meadows, V. S., Bitz, C. M., et al. 2013, AsBio, 13(8), 715.

Shields, A. L., Bitz, C. M., Meadows, V. S., et al. 2014, ApJL, 785, L9

Shields, A. L., Barnes, R., Agol, E., et al. 2016, AsBio, 16(6), 443

Tajika, E., 2007, EP&S, 59, 293

Urata, R. A. & Toon, O. B. 2013, *Icar*, 226, 229

Walker, J. C. G., Hays, P. B., & Kasting, J. F. 1981, JGR, 86, 9776

Way, M. J., Del Genio, A. D., Kiang, N. Y., et al. 2016, GRL, 43, **?**

Wolf, E. T. & Toon, O. B. 2013, AsBio, 13(7), 1

Wolf, E. T. & Toon, O. B. 2014, AsBio, 14(3), 241

Wolf, E. T. & Toon, O. B. 2014, GeoRL, 41, 167

Wolf, E. T. & Toon, O. B. 2015 JGRD, 120, 5775

Wordsworth, R. D., Forget, F., Selsis, F., et al. 2011, ApJL, 733, L48

Wordsworth, R., Forget, F., Millour, E., et al. 2013, Icar, 222, 1

Yang, J., Cowan, N., & Abbot, D. S., 2013, ApJL, 771, L45

Yang, J., Boue, G., Fabrycky, D. C., & Abbot, D. S. 2014, ApJL, 787, L2

Yang, J., Leconte, J., Wolf, E. T. et al. 2016, ApJ, 826, 222

Zhang, G. J. & McFarlane, N. A. 1995, Atmosphere-ocean 33(3), 407

Zsom, A., Seager, S., de Wit, J., & Stamenković, V. 2013, ApJ 778, 109

## Figure Captions

**Figure 1:** Empirical stellar spectra for F-, G-, and K-dwarf stars used in this study, normalized to a total flux of 1360 W $m^{-2}$.

**Figure 2**: The evolution of global mean surface temperatures (panels a,b,c) and climate sensitivity (panels d,e,f) for Earth-like planets around F-, G-, and K-dwarf main sequence stars as a function of relative stellar flux ($S/S_0$). The four stable climate regions are labeled and indicated by shaded regions in the top panels. Different color lines are associated with different simulation sets. Red lines indicate simulations starting from modern Earth conditions, under a positive solar forcing (*i.e.* warming). Green lines indicate simulations starting from modern Earth conditions, under a negative solar forcing (*i.e.* cooling). Finally, blue lines indicate simulations starting from a globally glaciated state, under a positive solar forcing. Numbers in the bottom panels mark peaks in climate sensitivity and thus represent specific climatic transitions; (1) from waterbelt to snowball, (2) from temperate to waterbelt, (3) from snowball to temperate, (4) temperate to moist greenhouse, and (5) towards a runaway greenhouse. The black dashed line marks the climate where diffusion limited water-loss could remove an Earth ocean of water within about 1 Gyr.

**Figure 3:** The model top temperature (a) and the model top water vapor volume mixing ratio as a function of mean surface temperature for planets around F-, G- and K-dwarf stars. Note the model top pressure is ~0.2 mb. Ocean loss timescales are calculated as the time for diffusion limited escape to remove an Earth ocean of water from the planet.

**Figure 4:** Circumstellar climate zones as a function of relative stellar flux for Earth-like planets at constant $CO_2$. The top panel assumes an initial state that is warm (i.e. liquid water covers the surface). The bottom panel assumes an initial state of a completely ice-covered planet.

**Figure 5.** The lifetime of habitable (solid lines) and temperate (dashed lines) climate zones driven by main sequence brightening, for warm (red) and cold (blue) initial conditions respectively, as a function of the stellar effective temperature. These values represent the maximum possible time for these phases of climates to exist, without invoking a draw-down of $CO_2$ to stabilize climate against continued warming.

**Figure 6:** Constraints on the inner edge (a), outer edge (b), and total habitable zone (c) determined from recent modeling studies. In all panels, tidally locked planets reside in the light yellow shaded region while planets in the blue shaded region can rotate rapidly. Solid lines and diamonds are used to mark constraints on the inner edge of the habitable zone. Dashed lines and triangles are used to mark constraints on the outer edge of the habitable zone.

**Table 1**

| Model configurations | Cold[1] | Warm[2] |
|---|---|---|
| Moist physics | CAM4 | CAM4 |
| Horizontal resolution | 2° x 2.5° | 4° x 5° |
| Vertical levels | 26 | 45 |
| Model top (mb) | 3 | 0.2 |
| Radiative transfer[3] | Native | Correlated-*k* |
| Longwave range (μm) | 5.0 – 1000 | 2.5 - 1000 |
| Shortwave range (μm) | 0.2 – 5.0 | 0.2 -12.2 |
| Continents | None | Present day Earth |
| Ocean heat transport | None | Present day Earth |
| Ocean albedo, visible/infrared | 0.07/0.06 | 0.07/0.06 |
| Sea ice albedo, visible/infrared | 0.67/0.3 | 0.68/0.3 |
| Snow albedo, visible/infrared | 0.8/0.68 | 0.91/0.63 |
| $CO_2$ (ppm) | 400 | 367 |

[1]Used for simulations into and out of snowball states, see Shields et al. (2013)
[2]Used for simulations of moist greenhouse states, see Wolf & Toon (2015)
[3]See Yang et al. (2016) for a comparison of these two codes.

**Table 1:** Summary of "cold" and "warm" model configurations. Each uses the same core atmosphere model, with the same model physics, except where noted above.

**Table 2**

| Star (type) | F | G | K |
|---|---|---|---|
| Surface Temperature (K) | 281.4 | 289.1 | 294.8 |
| 2-Meter Air Temperature (K) | 280.1 | 288.0 | 294.1 |
| Albedo, All-sky | 0.387 | 0.329 | 0.284 |
| Albedo, Clear-sky | 0.216 | 0.152 | 0.119 |
| Albedo, Cloud | 0.171 | 0.176 | 0.166 |
| Albedo, Surface | 0.143 | 0.114 | 0.104 |
| Albedo, Rayleigh | 0.073 | 0.038 | 0.015 |
| Greenhouse Effect, Total (K) | 35.3 | 37.6 | 39.5 |
| Greenhouse Effect, Clear-sky (K) | 25.0 | 28.5 | 31.3 |
| Greenhouse Effect, Cloud (K) | 10.3 | 9.1 | 8.2 |
| Water Vapor Column (Kg $m^{-2}$) | 15.7 | 26.5 | 41.2 |
| Cloud Water Column (Kg $m^{-2}$) | 0.093 | 0.115 | 0.129 |
| Cloud Ice Column (Kg $m^{-2}$) | 0.018 | 0.015 | 0.013 |
| Cloud Fraction, Total (%) | 69.5 | 66.3 | 64.3 |
| Cloud Fraction, Low (%) | 35.7 | 34.8 | 37.7 |
| Cloud Fraction, Middle (%) | 29.9 | 27.0 | 23.3 |
| Cloud Fraction, High (%) | 49.6 | 47.2 | 43.4 |
| Sea Ice Fraction Relative to Ocean (%) | 16.1 | 7.5 | 3.7 |
| Snow Depth (m) | 0.081 | 0.042 | 0.021 |

**Table 2:** Global and annual mean quantities from control simulations using the warm configuration, which includes ocean heat transport identical to the modern Earth.

**Table 3**

| Differentiations of Circumstellar Climate | a | b | c |
|---|---|---|---|
| Water-loss 1 Gyr ($T_s$ = 355 K)[1] | 1.19645 | $1.39815\times10^{-4}$ | $3.12706\times10^{-8}$ |
| Temperate to Moist Greenhouse Transition | 1.11892 | $8.48102\times10^{-5}$ | $3.89303\times10^{-8}$ |
| Present Day Earth Conditions ($T_s$ = 289 K)[2] | 1.00014 | $6.81156\times10^{-5}$ | $2.12922\times10^{-8}$ |
| Biological Heat Stress ($T_s$ = 300 K) | 1.06666 | $8.43240\times10^{-5}$ | $2.61308\times10^{-8}$ |
| Temperate to Waterbelt Transition[2] | 0.96011 | $5.54441\times10^{-5}$ | $7.18032\times10^{-9}$ |
| Waterbelt to Snowball Transition[2] | 0.92515 | $7.27318\times10^{-5}$ | $9.82310\times10^{-10}$ |
| Snowball to Temperate Transition[3] | 1.05521 | $1.07307\times10^{-4}$ | $-1.14135\times10^{-8}$ |

[1]The water-loss limit is the nominal inner edge of the habitable zone.
[2]Only accessible from warm start conditions
[3]Only realizable from cold start conditions

**Table 3:**. Coefficients *a*, *b*, and *c* to be used in equation 1 to calculate circumstellar climates zones in units of relative stellar flux for rapidly rotating Earth-like planets. These parameterizations were fit to our model calculations. Valid for stars with 4900 K ≥ $T_{eff}$ ≥ 6600 K.

## Figure 1

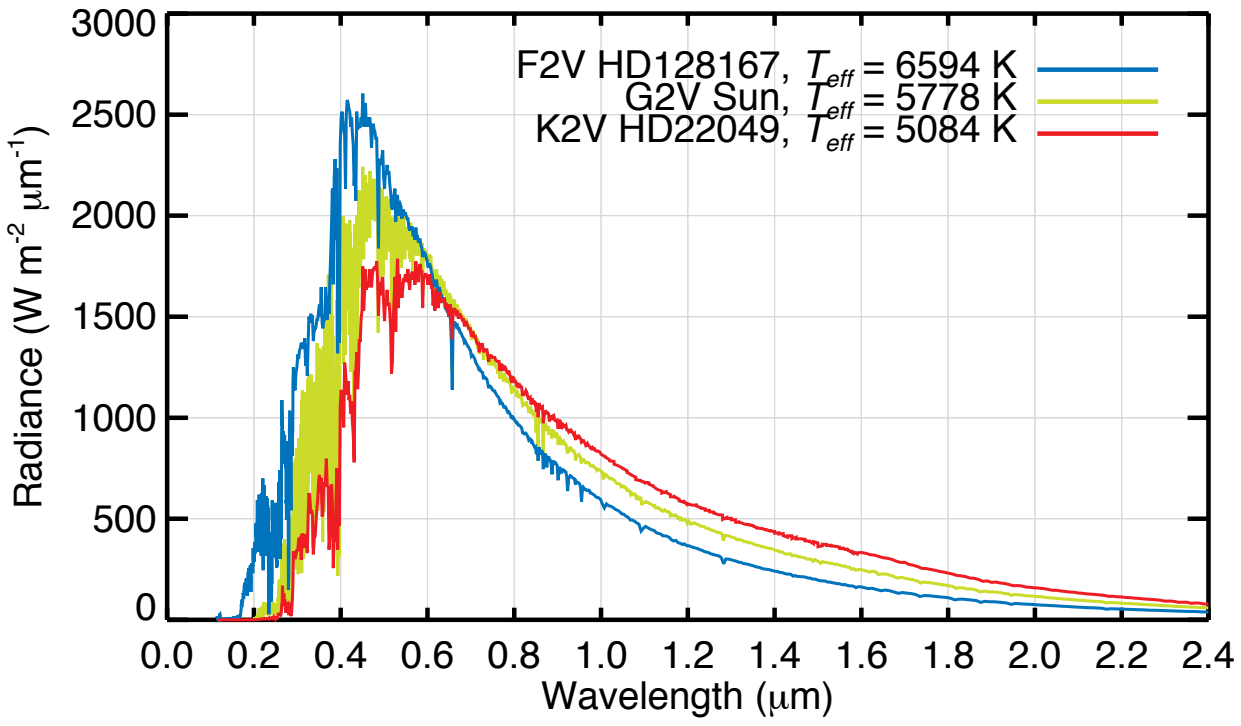

F2V HD128167, $T_{eff}$ = 6594 K
G2V Sun, $T_{eff}$ = 5778 K
K2V HD22049, $T_{eff}$ = 5084 K
Radiance (W m$^{-2}$ μm$^{-1}$)
3000
2500
2000
1500
1000
500
0
0.0 0.2 0.4 0.6 0.8 1.0 1.2 1.4 1.6 1.8 2.0 2.2 2.4
Wavelength (μm)

# Figure 2

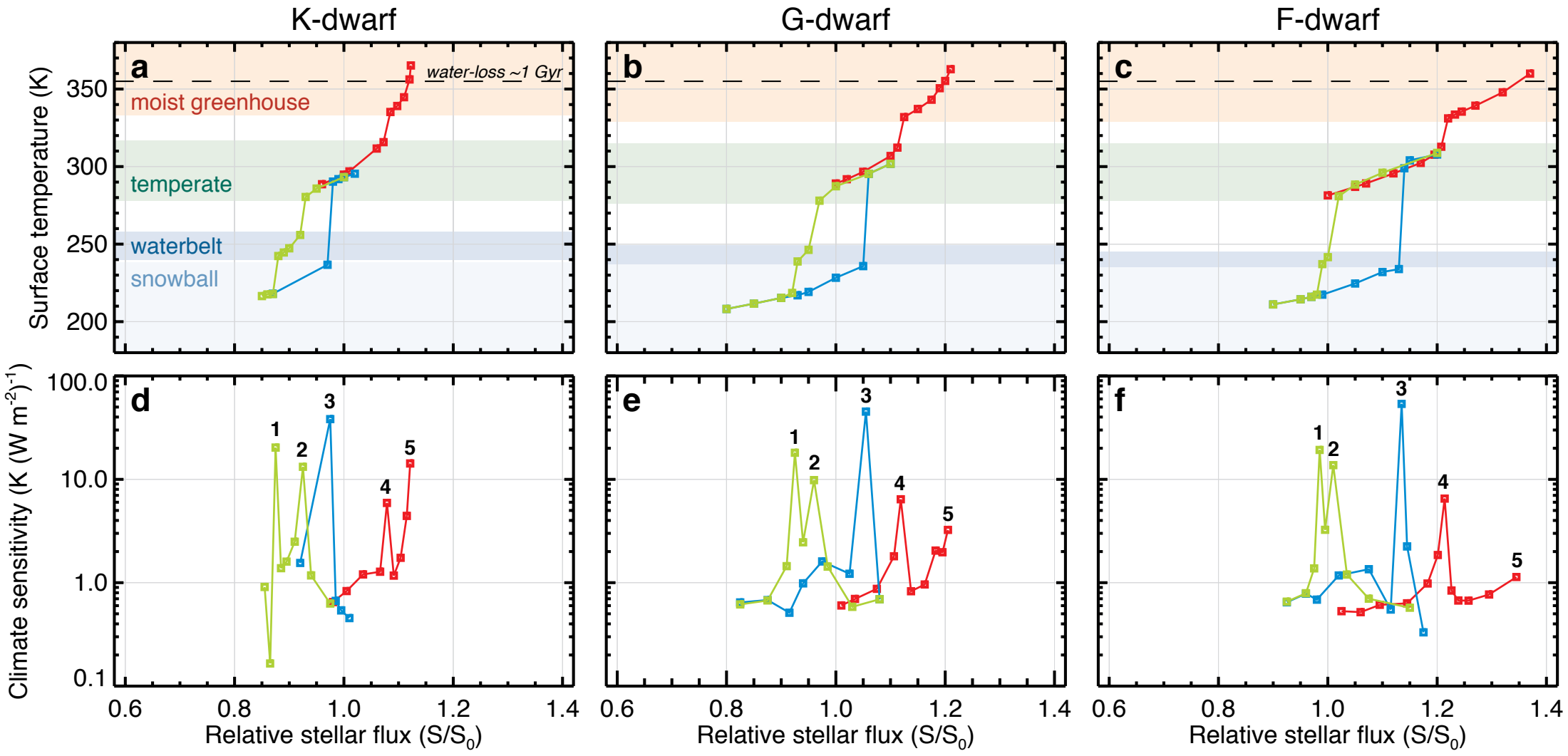

# Figure 3

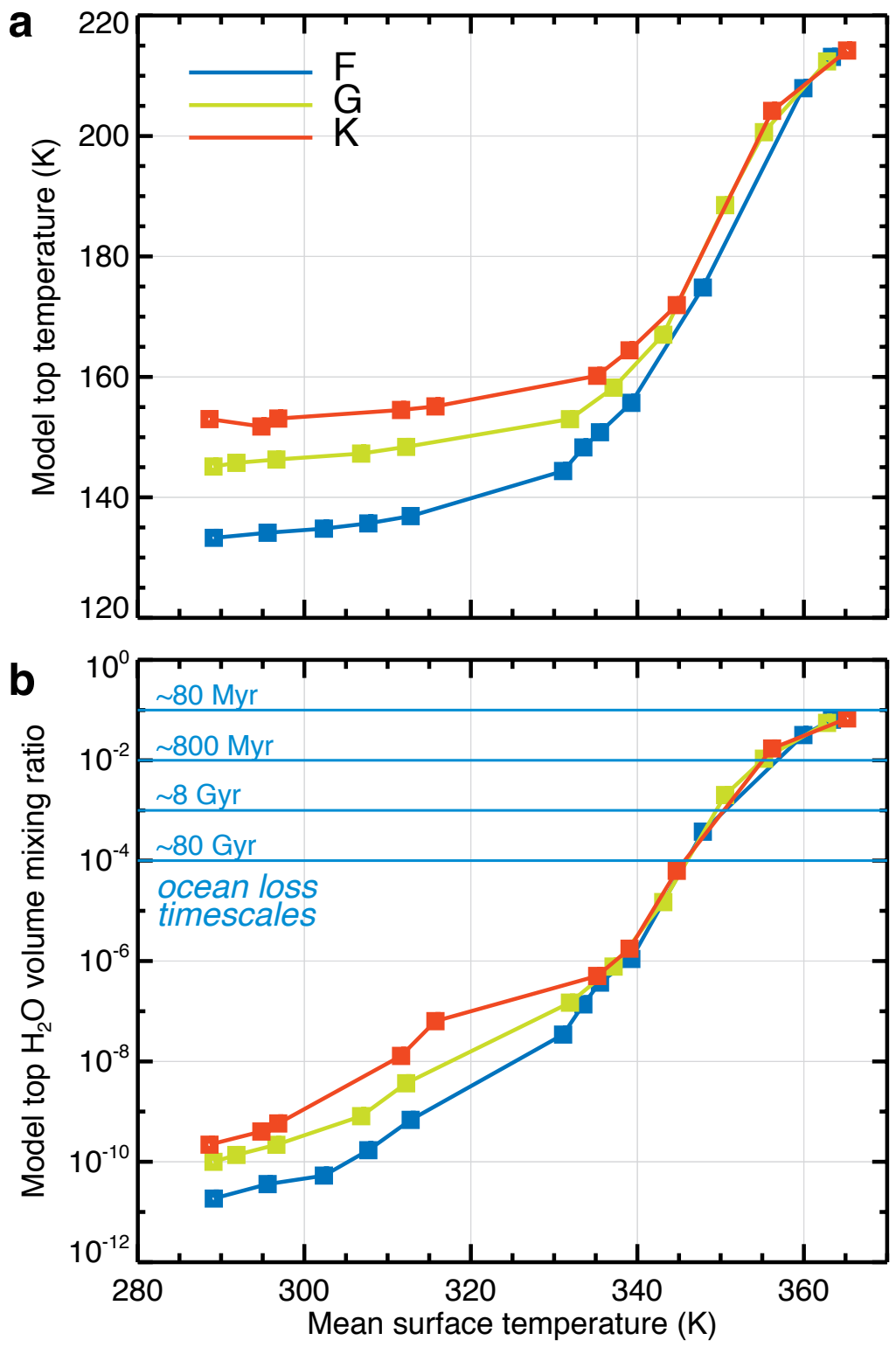

**Figure 4**

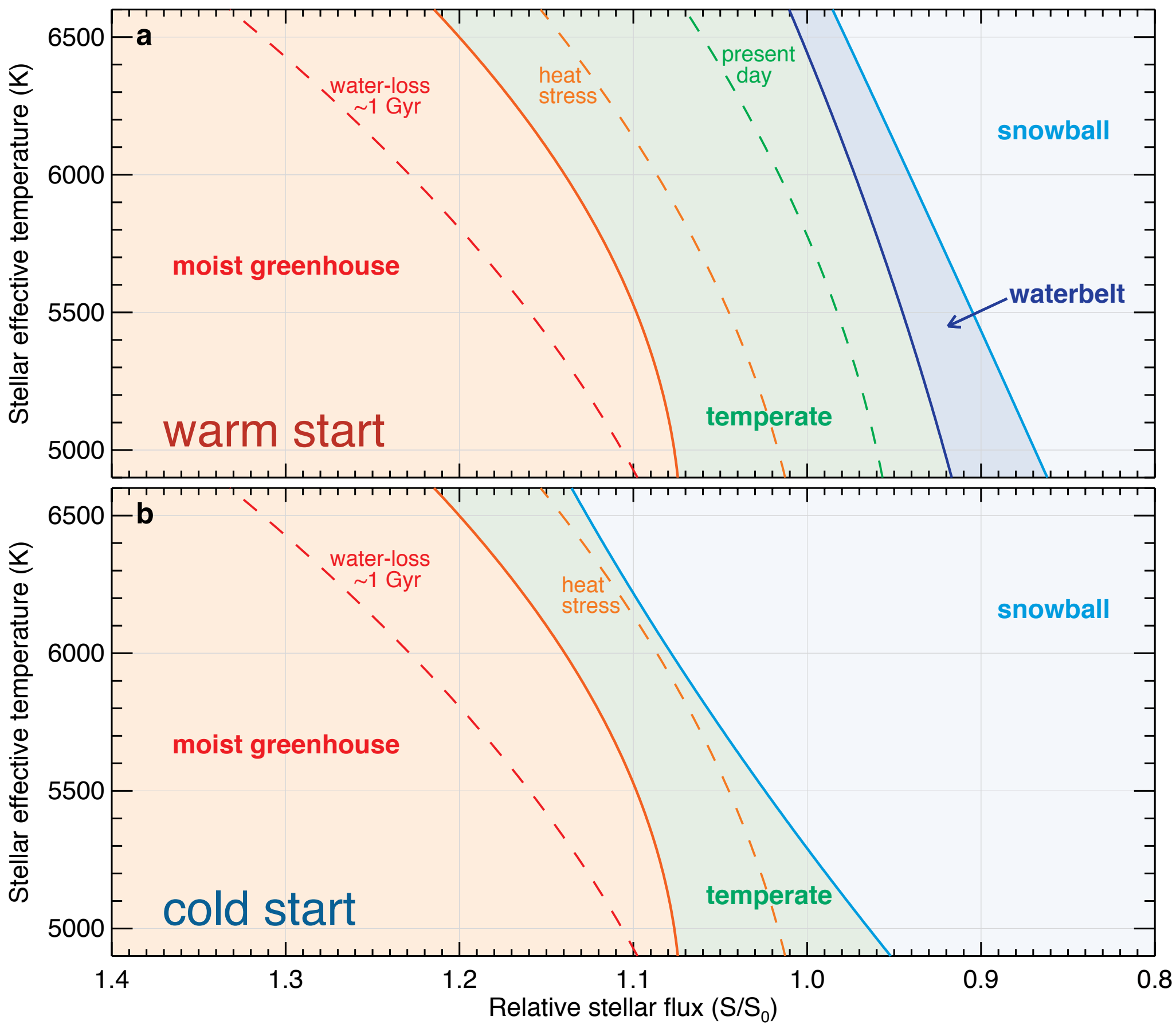

# Figure 5

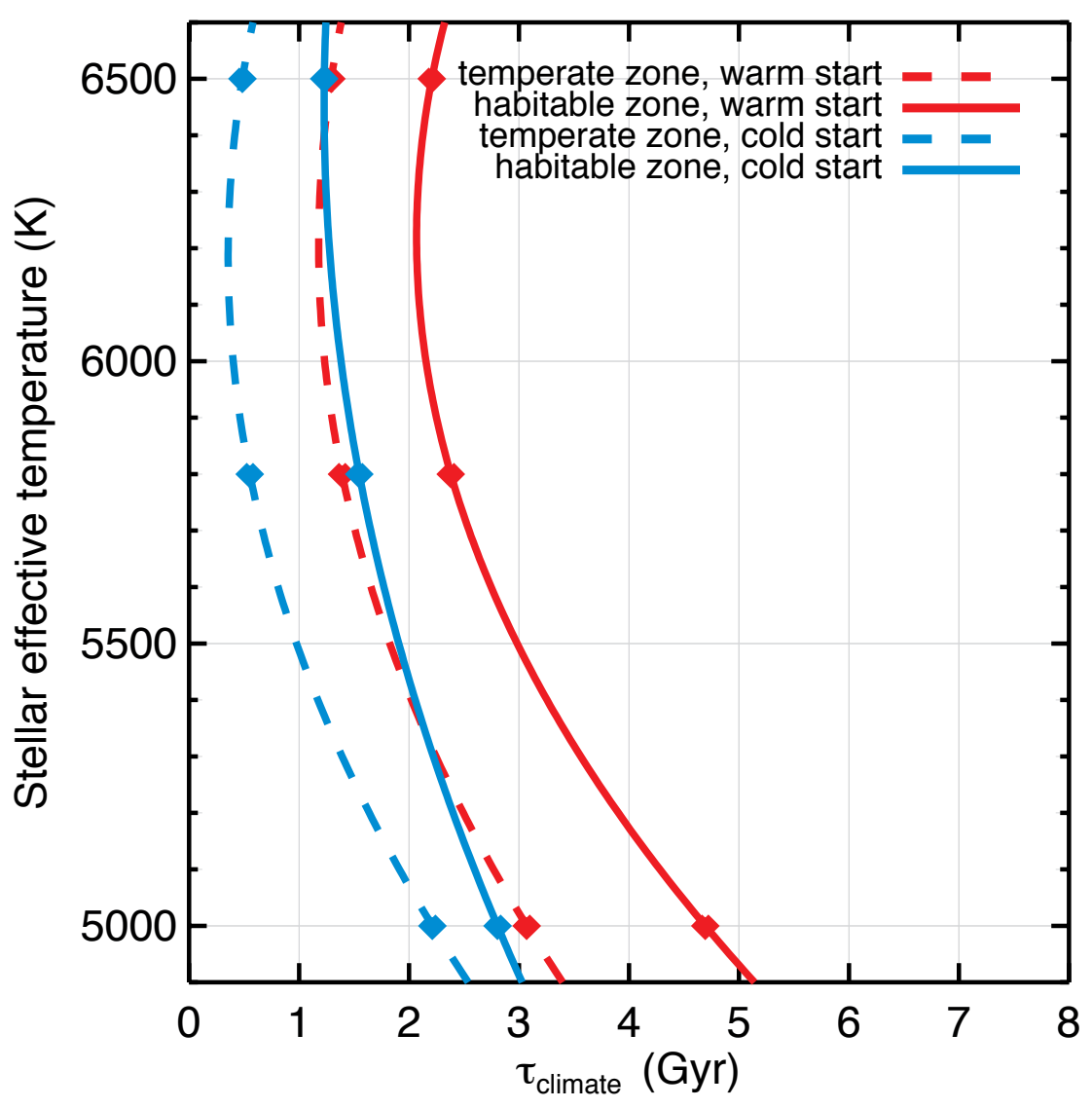

temperate zone, warm start
habitable zone, warm start
temperate zone, cold start
habitable zone, cold start
Stellar effective temperature (K)
6500
6000
5500
5000
0
1
2
3
4
5
6
7
8
τclimate (Gyr)

**Figure 6**

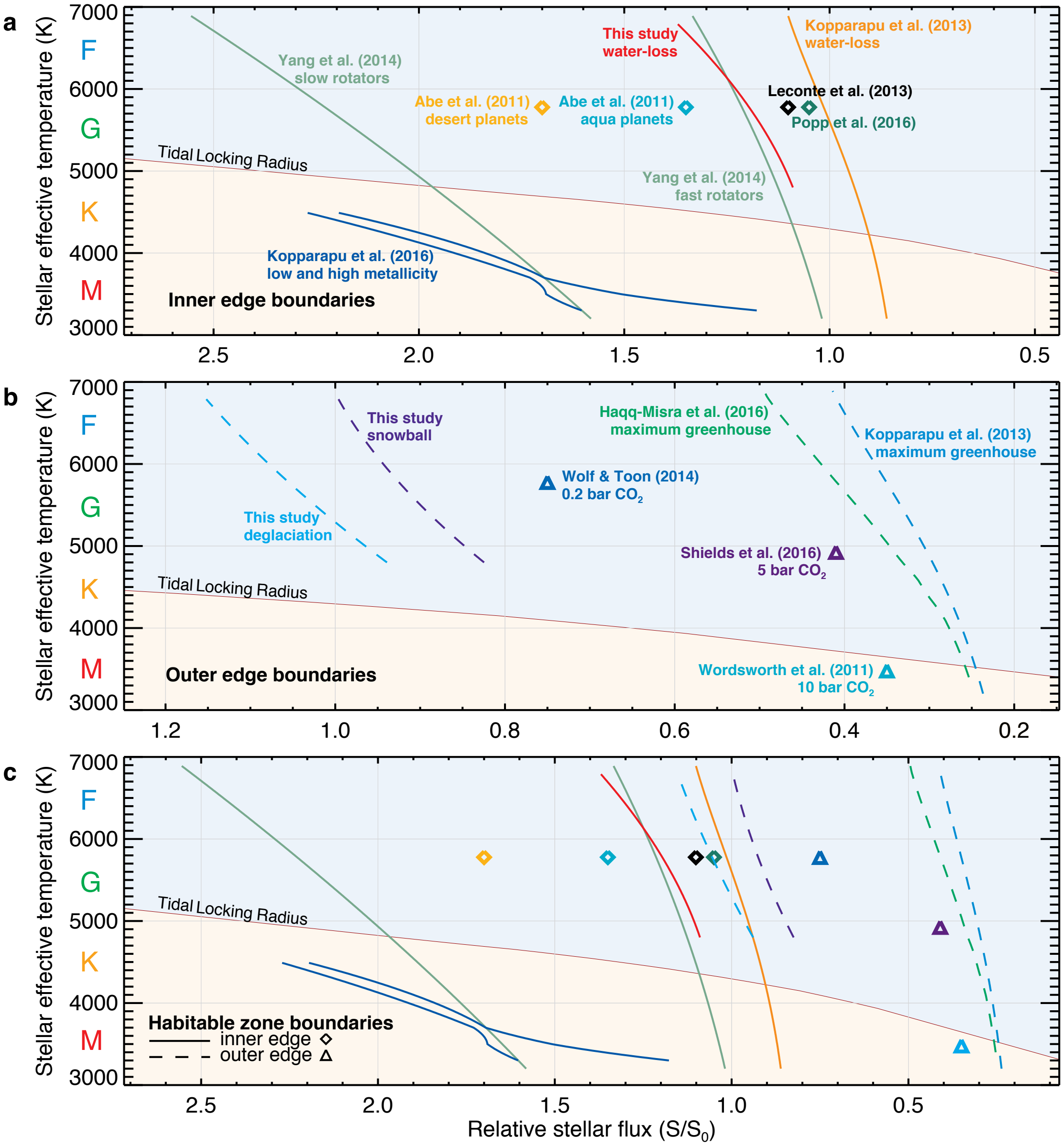

a
Inner edge boundaries
This study water-loss
Kopparapu et al. (2013) water-loss
Yang et al. (2014) slow rotators
Abe et al. (2011) desert planets
Abe et al. (2011) aqua planets
Leconte et al. (2013)
Popp et al. (2016)
Yang et al. (2014) fast rotators
Tidal Locking Radius
Kopparapu et al. (2016) low and high metallicity
b
Outer edge boundaries
This study snowball
Haqq-Misra et al. (2016) maximum greenhouse
Kopparapu et al. (2013) maximum greenhouse
Wolf & Toon (2014) 0.2 bar $CO_2$
This study deglaciation
Shields et al. (2016) 5 bar $CO_2$
Wordsworth et al. (2011) 10 bar $CO_2$
c
Habitable zone boundaries
inner edge
outer edge
Stellar effective temperature (K)
F
G
K
M
Relative stellar flux ($S/S_0$)